\pdfoutput=1
\documentclass[12pt]{prop}
\usepackage{apjfonts}
\usepackage{overcite}
\usepackage{graphicx}

\def\chandra    {\emph{Chandra}}
\def\xmm        {\emph{XMM}}
\def\astroh     {\emph{Astro-H}}

\def\wmap       {\emph{WMAP}}

\def\ixo     {\emph{IXO}}
\def\generx     {\emph{Generation-X}}
\def\genx     {\emph{Gen-X}}
\def\fermi    {\emph{Fermi}}
\def\lofar    {\emph{LOFAR}}
\def\lwa    {\emph{LWA}}
\def\ska    {\emph{SKA}}
\def\gmrt    {\emph{GMRT}}

\def\deg        {$^{\circ}$}
\def\lesssim{\mathrel{\hbox{\rlap{\hbox{\lower4pt\hbox{$\sim$}}}\hbox{$<$}}}}
\def\gtrsim{\mathrel{\hbox{\rlap{\hbox{\lower4pt\hbox{$\sim$}}}\hbox{$>$}}}}
\def\lax{\lesssim}

\def\bi{\bfseries\itshape}

\renewcommand{\textfloatsep}{4mm}

\begin{document}
\sloppy

\pagestyle{empty}
\begin{centering}

\vspace{1cm}
{\Large DIFFUSE BARYONIC MATTER BEYOND 2020}

\vspace{1cm}
{\sc M. Markevitch, F. Nicastro, P. Nulsen, E. Rasia, 

A. Vikhlinin, A. Kravtsov, W. Forman, G. Brunetti, C. Sarazin, 

M. Elvis, G. Fabbiano, A. Hornschemeier, R. Brissenden,

and the Generation-X Team}

\vspace{1cm}
White Paper for the NRC Astro-2010 Decadal Survey

\vspace{4cm}
%%%%%%%%%%%%%%%%%%%%%%%%%%%%%%%%%%%%%%%%%%%%%%
\includegraphics[width=0.65\textwidth]%
{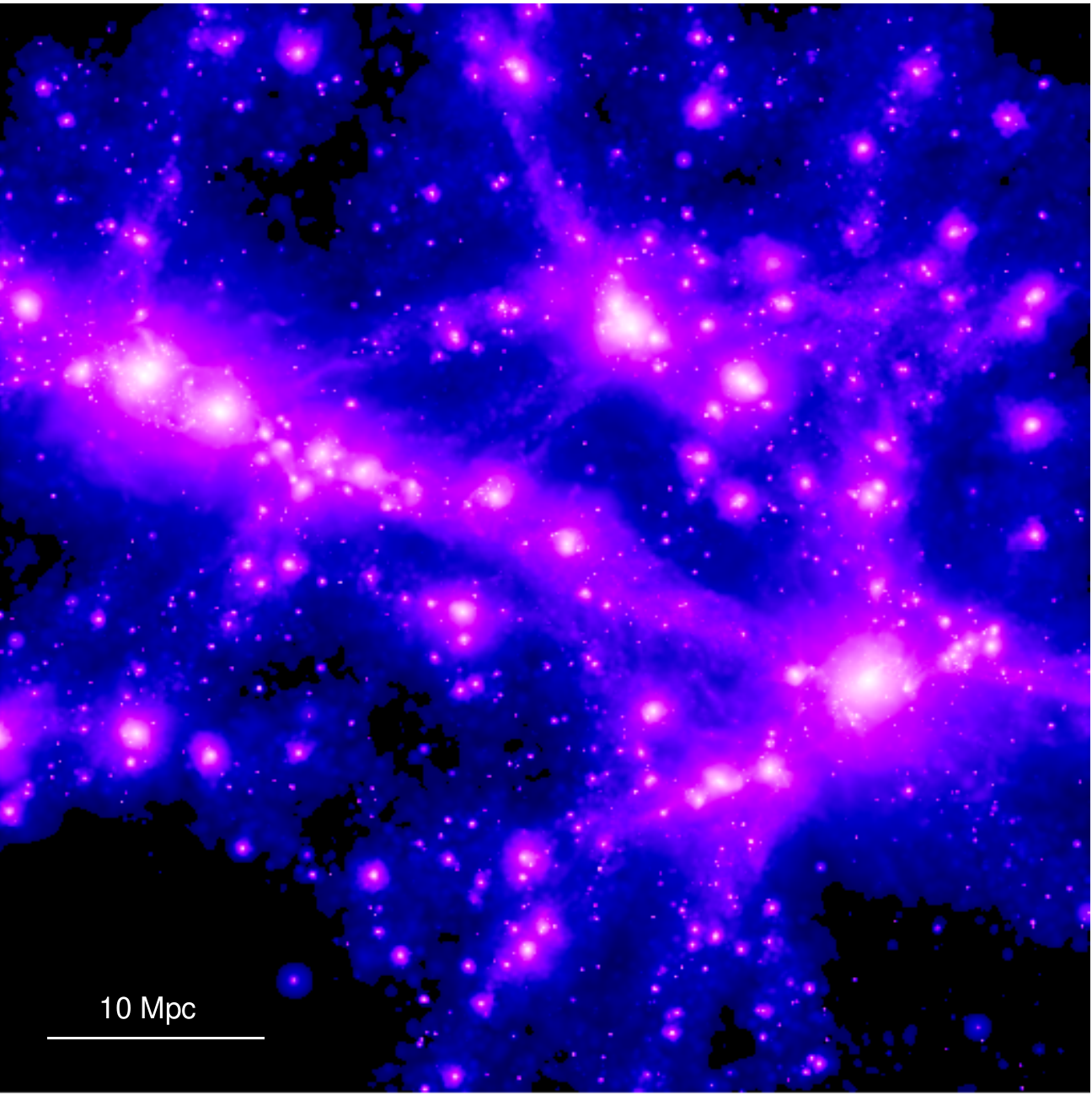}
%%%%%%%%%%%%%%%%%%%%%%%%%%%%%%%%%%%%%%%%%%%%%%

\vfill
February 15, 2009

\end{centering}

\clearpage
\twocolumn

\pagestyle{plain}
\setcounter{page}{1}

%%%%%%%%%%%%%%%%%%%%%%%%%%%%%%%%%%%%%%%%%%%%%%%%%%%%%%%%%%%%%%%%%%%%%%%%%
\section*{ABSTRACT}

The hot, diffuse gas that fills the largest overdense structures in the
Universe --- clusters of galaxies and a web of giant filaments connecting
them --- provides us with tools to address a wide array of fundamental
astrophysical and cosmological questions via observations in the X-ray band.
Clusters are sensitive cosmological probes. To utilize their full potential
for precision cosmology in the following decades, we must precisely
understand their physics --- from their cool cores stirred by jets produced
by the central supermassive black hole (itself fed by inflow of intracluster
gas), to their outskirts, where the infall of intergalactic medium (IGM)
drives shocks and accelerates cosmic rays.  Beyond the cluster confines lies
the virtually unexplored warm IGM, believed to contain most of the baryonic
matter in the present-day Universe.  As a depository of all the matter ever
ejected from galaxies, it carries unique information on the history of
energy and metal production in the Universe.  Currently planned major
observatories, such as \astroh\ and \ixo, will make deep inroads into these
areas, but to see the most interesting parts of the picture will require an
almost science-fiction-grade facility with tens of m$^2$ of effective area,
subarcsecond angular resolution, a matching imaging calorimeter and a super
high-dispersion spectrograph, such as \generx.

%%%%%%%%%%%%%%%%%%%%%%%%%%%%%%%%%%%%%%%%%%%%%%%%%%%%%%%%%%%%%%%%%%%%%%%%%
\section{OVERVIEW AND RECENT ADVANCES}

Most of the visible matter in the Universe is in the form of diffuse gas
that fills dips and valleys of the Universe's gravitational potential. It is
heated to $T\sim 10^5-10^8$ K by shocks generated by the growth of Large
Scale Structure (LSS).  At present, we can study only the hottest and
densest phase of this matter,
found in central regions of galaxy clusters ($r<r_{500}-r_{200}$%
\footnote{Radii of the average overdensity of 500 and 200 above the
  critical density of the Universe}%
), which comprises only a small fraction of the total.  The gas within these
regions is close to hydrostatic, and its X-ray observables can be used to
estimate the cluster total (dark matter dominated) masses
\cite{bahcallsarazin77,sarazin88book}, providing the basis for sensitive
cosmological tests (\S\ref{sec:cosmol}).

There are deviations from hydrostatic equilibrium, however, observed in
clusters undergoing a growth event and in many cool-core clusters, where
jets and relativistic plasma from the central supermassive black hole stir
the gas.  Enormous progress in understanding these phenomena has been made
in the past decade with the advent of powerful X-ray imaging spectrographs
such as \xmm\ and \chandra.  \xmm\ has determined that radiative cooling of
the dense cores must be compensated by some heating mechanism
\cite{petersonfabian06rev}.  \chandra\ provided a leap in angular resolution
that has led to the discovery of the ubiquitous AGN-blown, radio-filled
bubbles in most cool cores \cite{mcnamara00hyda,blanton01a2052,forman05m87}.
It called into question the old paradigm that gravity is the only important
source of thermal energy for the intracluster medium (ICM) --- apparently,
AGN can inject as much mechanical energy into the core gas as the gas loses
via radiative cooling\cite{bohringer02,chur02agnmech,mcnamaranulsen07cfrev}.
Precisely how this injection works, and how much of the cluster volume is
affected, is unclear (\S\ref{sec:bubbles-hiz}).

Merging clusters revealed a wealth of gas motion-related phenomena ---
subcluster infall, shocks, ``cold fronts'', cool core sloshing, ram pressure
stripping --- all deduced indirectly using \chandra's high-resolution
imaging and modest spectral information \cite{m00a2142, vikhl01a3667cf,
  m02textbook, mazz03-2a0335, mv07rev}.  These phenomena await an imaging
calorimeter, which will measure the gas velocities directly
(\S\ref{sec:vels}). One of the surprises was the ubiquity, sharpness and
symmetry of ``cold fronts'' and the stability of AGN bubbles, indicating
that mixing and instabilities in the ICM are suppressed by some unexpected
microphysical properties of the intracluster plasma.

But perhaps the most interesting regions of clusters lie beyond the reach of
the current X-ray instruments, because of their extremely low surface
brightness. These are regions where the intergalactic medium (IGM) that
flows along giant filaments of the Cosmic Web meets the intracluster gas.
Physical processes in those regions hold the key to a number of important
astrophysical and cosmological questions. Most of the baryonic matter in the
present Universe is located still further into the low-density regions of
the Cosmic Web. It is accessible only via FUV and X-ray line absorption
studies that require very large collecting areas and very high spectral
resolution (\S\ref{sec:whim-abs}). So far only a few tentative detections of
such absorption lines have been reported.%
\cite{nicastro05whima,nicastro05whimn,danforthshull08,tripp08,richter06}

In this paper, we consider which fundamental questions in this field will
remain {\em beyond}\/ the capabilities of the near-term, trailblazing X-ray
facilities such as \astroh\ and \ixo. To answer them, technological leaps
such as those proposed in the \generx\ mission concept (White Paper by
Schwartz et al.) are needed during the coming decade.

%%%%%%%%%%%%%%%%%%%%%%%%%%%%%%%%%%%%%%%%%%%%%%%%%%%%%%%%%%%%%%%%%%%%%%%%%%%%%%
\section{COSMOLOGY WITH GALAXY CLUSTERS}
\label{sec:cosmol}

Clusters are the most massive virialized structures in the Universe, which
makes them sensitive cosmological probes. Two avenues have been actively
pursued --- the {\em growth of structure}\/ test based on the evolution of
the cluster mass function \cite{press74,henry97evol,vikhl08cosmol}, and the
{\em geometric}\/ test based on the cluster baryon fraction
\cite{allen08cosmol}. At the present accuracy, the results strongly support
the need for Dark Energy and are uniquely complementary to other studies
(CMB, SNe, galactic surveys).  Given the breadth of survey and cosmology
missions planned for the near future, it is impossible to foresee what
parameters of the current cosmological model will still be of interest 1-2
decades from now. In a model-independent way, clusters can uniquely trace
the growth of LSS from $z\sim 1.5-2$ to the present.  Its precise behavior
depends on the nature of Dark Energy (WP by Vikhlinin et al).  High-$z$
clusters for these studies will be found in large numbers by the forthcoming
SZ and/or X-ray survey missions, but to estimate cluster masses, an angular
resolution and collecting area of \ixo\ or, for higher $z$, \genx\ will be
required.

Statistical uncertainties arising from small sample sizes remain the
dominant uncertainty in current studies. It is increasingly clear, however,
that any precision cosmological studies in the next 1-2 decades, which will
take advantage of many hundreds of clusters, will need cluster mass
estimates with a systematic accuracy of $\sim 1-3\%$ (this is not the
accuracy of {\em individual}\/ cluster masses, but biases for sample
averages). A promising approach is to use future sensitive weak lensing mass
measurements to calibrate a well-behaved combination of X-ray observables
and use it as a mass proxy (WP by Vikhlinin et al). While this can be done
on a purely statistical basis, experience tells us that understanding and
adequate numerical modeling of the underlying physics is key to obtaining
robust constraints. Hence, {\bi cluster precision cosmology in the next
  decades will look more like cluster physics.}\/ The necessary new physical
knowledge will come from combining gravitational lensing, SZ, low-frequency
radio, and gamma-ray observations, all of which will come of age in the next
decade, with vastly improved X-ray capabilities.  For example, the currently
identifiable questions that may affect cosmological constraints are:

$\bullet$ Why is cluster $f_{\rm bar}$, the fraction of baryons (gas +
stars) in the total mass, slightly lower than the Universal average?

$\bullet$ How relaxed are ``relaxed'' clusters? What is the fraction of
turbulent and nonthermal pressure components (cosmic rays, magnetic fields)
in the total gas pressure?

%%%%%%%%%%%%%%%%%%%%%%%%%%%%%%%%%%%%%%%%%%%%%%%%%%%%%%%%%%%%%%%%%%%%%%%%%%%
\subsection{``Missing baryons'' inside clusters}
\label{sec:bubbles-hiz}

The baryon fraction $f_{\rm bar}$ derived from the X-ray data within
$r<r_{2500}-r_{500}$, is too low by $\sim 30$\% compared to the Universal
value from \wmap\cite{ettori03missbar,vikhl06mass}. This is lower than
expected if gravity and LSS shocks were the only significant heat source for
the ICM. As this assumption is basic for cluster cosmology studies, we need
to map this discrepancy to greater radii, investigate its evolution with
$z$, explain its cause and eventually include adequate modeling into numeric
codes (for a discussion see WP by Kravtsov et al).  Among the possible
explanations are (a) some cosmological parameters involved in the cluster
$f_{\rm bar}$ derivation, such as $H_0$, or even the universal $f_{\rm bar}$
itself, are incorrect; (b) a systematic overestimate of cluster total
masses; (c) large mass in unseen stars into which the ICM has been
converted, or a gross underestimate of the stellar $M/L$ ratio; and (d)
energy injection from supernovae or AGN before or during cluster formation.
As suggested by the recent discovery of half-Mpc size ``ghost cavities'' in
some clusters \cite{mcnamara05bigcavity,wise07hyda}, feedback from the
central AGN can affect ICM over a much greater cluster volume than thought
\cite{mcnamaranulsen07cfrev}. The first three possibilities will be
addressed by forthcoming cosmology missions and sensitive gravitational
lensing and optical measurements.  The last possibility will be probed in
the course of a fundamental study of the growth of supermassive black holes
in the cluster centers over the cosmic time.

\subsection{Cosmological history of AGN feedback}

X-ray cavities, filled with relativistic plasma, serve as ``calorimeters''
of the total power emitted by the cluster central black hole as it accretes
the intracluster gas\cite{chur02agnmech,mcnamaranulsen07cfrev}. To see the
emergence and evolution of these monster black holes will involve surveying
for AGN bubbles in the cores of clusters at $z=1-2$.  Because the CMB energy
density grows as $(1+z)^4$, at some redshift we may see the young X-ray
cavities fill up and become bright spots, because of the increased Inverse
Compton emission from the cosmic rays inside the bubbles.  Combined with
radio synchrotron data, this would open a unique window into the content and
energetics of cosmic rays (currently completely uncertain) and their effect
on the ICM.

These high $z$\/ observations will pose a technical challenge, because the
X-ray brightness contrast of the subtle ``ghost bubbles'' is very low, while
young bubbles are very small. An instrument with \chandra\ or better angular
resolution but a much greater effective area is required.  These studies
will be synergistic with sensitive low-frequency radio observations, e.g.,
\lofar, \lwa, \ska.

%%%%%%%%%%%%%%%%%%%%%%%%%%%%%%%%%%%%%%%%%%%%%%
\begin{figure*}
\vspace*{5mm}
\center
\includegraphics[width=0.45\textwidth]%
{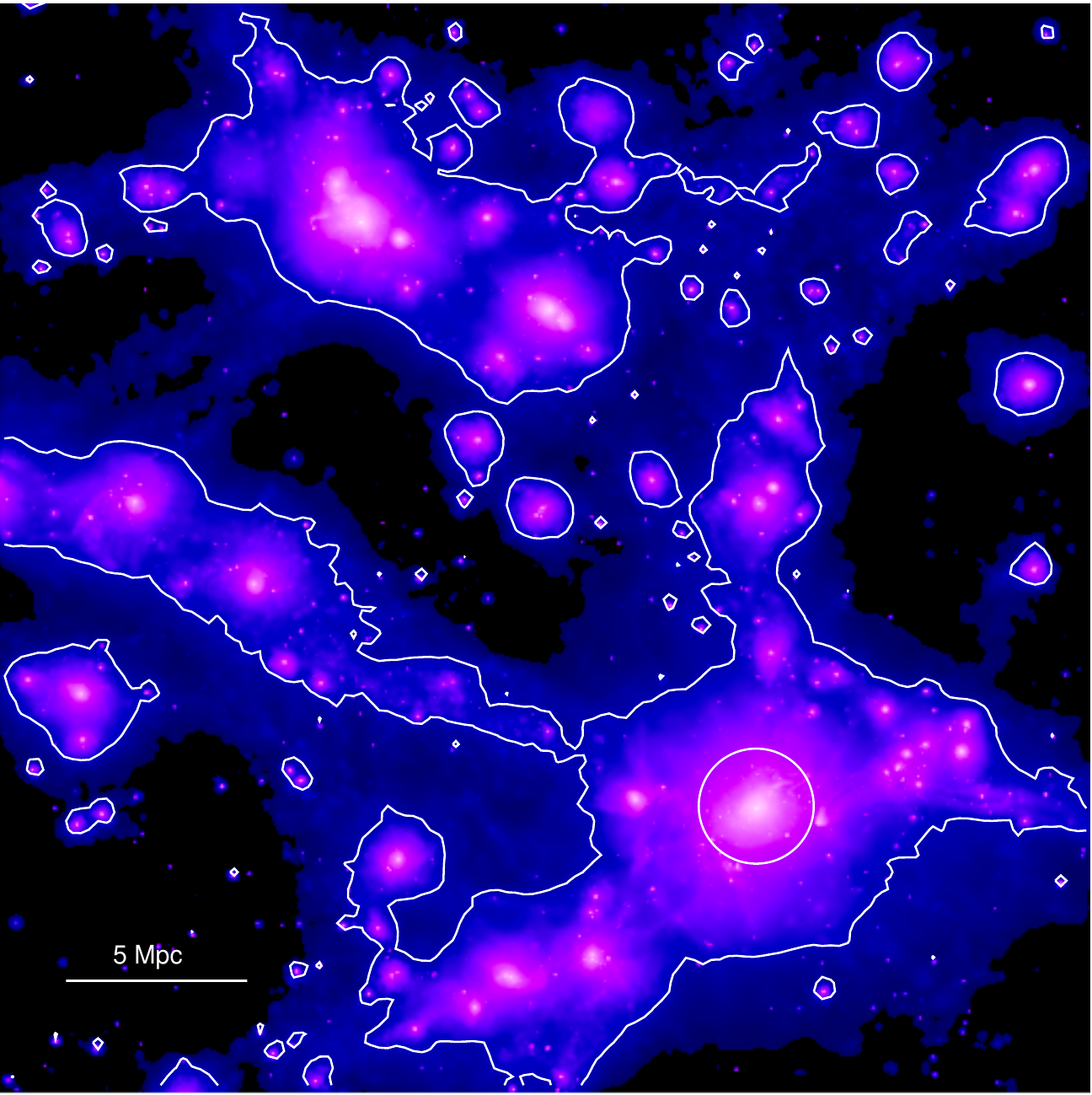}
\hspace{5mm}
\includegraphics[width=0.45\textwidth]%
{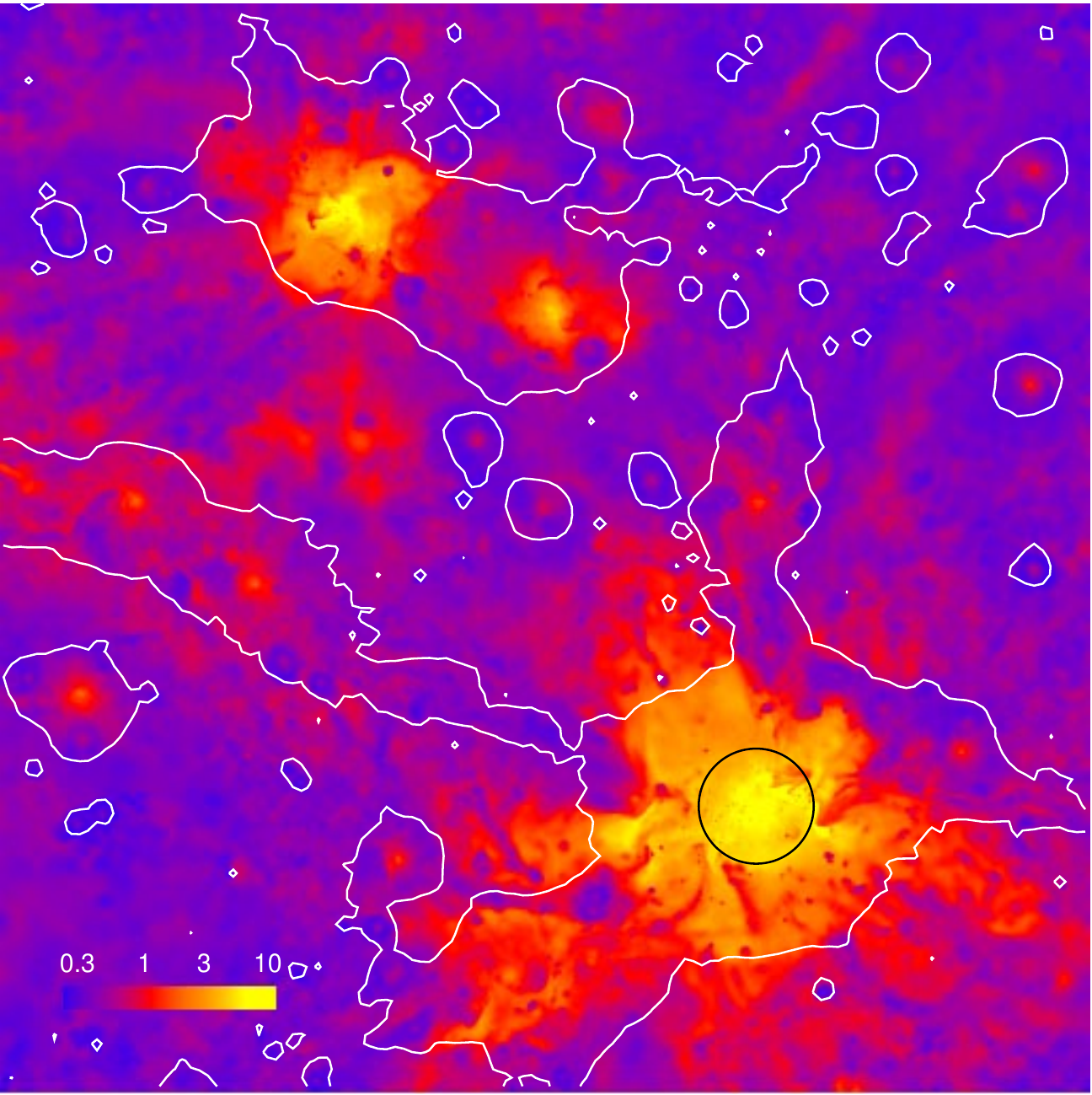}

\caption{High-resolution simulations of a $30\times 30$ Mpc region at an
  intersection of several filaments of the Cosmic Web. {\em Left:}\/ X-ray
  brightness in 0.5--2 keV band. {\em Right:}\/ gas temperature. The color
  scale bar is in keV. In both panels, the circle shows the $r_{500}$ region
  for the most massive cluster that is accessible for current telescopes,
  and the white contour shows the X-ray brightness 30 times below that at
  $r_{200}$, encompassing most of the interesting physics (simulations from
  ref.\ \citen{dolag06filam} and E. Rasia)}

\label{fig:elenamaps}
\end{figure*}
%%%%%%%%%%%%%%%%%%%%%%%%%%%%%%%%%%%%%%%%%%%%%%

%%%%%%%%%%%%%%%%%%%%%%%%%%%%%%%%%%%%%%%%%%%%%%%%%%%%%%%%%%%%%%%%%%%%%%%%%%%
\subsection{Cluster outskirts}
\label{sec:outskirts}

A solution to the cluster missing baryons problem, and clues to the cluster
nonthermal energy content, may well be found in the cluster outskirts,
between the currently observable region within $r<r_{500}$ from the cluster
center and the infall shock region around $r\sim 2 r_{200} \sim 3-5$ Mpc,
where the IGM, flowing along the Cosmic Web, meets the cluster.  Fig.\ 
\ref{fig:elenamaps} shows a simulated X-ray image of a supercluster; a
circle marks the central $r_{500}$ region of a massive cluster and the
brightness contour corresponds to $r\sim 2 r_{200}$.  The area between them
is clearly where most of the action is --- shocks and turbulence are
generated, small infalling gas halos are stripped from their collisionless
dark matter hosts, the near-pristine IGM is mixed (or not) with metal-rich
halos and heated to $T>1$ keV.  By combining weak lensing mapping of the
total mass and X-ray + SZ mapping of the gas, we may find that the gas
fraction does indeed approach the Universal value, or find clues why it does
not. We are also sure to discover a lot more.

Physical processes occurring in this region determine the matter and energy
content of clusters. For example, infall shocks, with $M\sim 10$, should be
efficient accelerators of cosmic rays
\cite{ryu03cosmolshocks,hoeftbruggen07,blasi07accelrev}, though much less
efficient than strong shocks in supernovae \cite{kangjones07acceleffic}.
Data on these shocks will be invaluable for studying the CR acceleration
mechanisms, with implications throughout the astrophysics.  These cosmic rays
should then be advected into clusters, lurk there for
gigayears\cite{sarazin99fossilcr} and provide seeds for reacceleration by
merger shocks and turbulence, giving rise to cluster radio
halos\cite{brunetti08natur}.  The cosmic ray content of clusters, especially
the relativistic protons, is presently unknown; \fermi\ and Cherenkov
telescopes are expected to test the most extreme scenarios.  Relativistic
electrons accelerated at those far-flung shock fronts should be seen at low
radio frequencies by, e.g., \gmrt, \lofar, \lwa, and eventually \ska\ (see WP
by Rudnick et al). This is the area where combining radio and X-ray data
would be extremely synergistic --- the SZ providing thermal pressure, the
low-frequency radio exploring the nonthermal phenomena, and the X-ray
yielding the gas density and temperature, along with Mach number
determination for any shocks discovered.

The gas in the shock regions will have elemental abundances representative
of the intergalactic space, but be compressed and heated to an observable
state. By comparing the abundances in the shocked gas to those in the
intracluster gas, we can learn about chemical history of the Universe.

Because the X-ray surface brightness in these regions is very low, this is
uncharted territory. In addition to large effective area and low detector
background, it requires one to remove the CXB, whose brightness is much
higher. \chandra\ or better angular resolution is needed to resolve its
point source component, and a 1-2 eV spectral resolution to resolve the
Galactic diffuse foreground, consisting mostly of nonredshifted emission
lines.  A calorimeter will also be needed to study the IGM emission lines
that dominate at $T\lax 1$ keV.

Studies of X-ray emission from cluster outskirts can be uniquely combined
with absorption lines in high-dispersion spectra of the background quasars
(see \S\ref{sec:clust-abs} below). \genx\ will have the sensitivity to probe
to the white contour in Fig.\ \ref{fig:elenamaps}, both in emission and
absorption, unlike any of the planned missions.

%%%%%%%%%%%%%%%%%%%%%%%%%%%%%%%%%%%%%%%%%%%%%%%%%%%%%%%%%%%%%%%%%%%%%%%%%%%
\subsection{Particle cosmology}

Some of the presently attractive Dark Matter candidates can be discovered by
X-ray observations before they are discovered in the laboratory. For
example, sterile neutrinos are expected to decay and produce an X-ray
emission line\cite{dodelsonwidrow94}. Recent observations of large DM
concentrations such as clusters and our Galaxy constrained the simple sterile
neutrino models (e.g., \citen{abazajian07,boyarsky08}), but a discovery
space from the current upper limits down to zero is wide open for a
sensitive imaging calorimeter.

%%%%%%%%%%%%%%%%%%%%%%%%%%%%%%%%%%%%%%%%%%%%%%
\begin{figure*}
\vspace*{5mm}
\center
\includegraphics[width=0.325\textwidth]%
{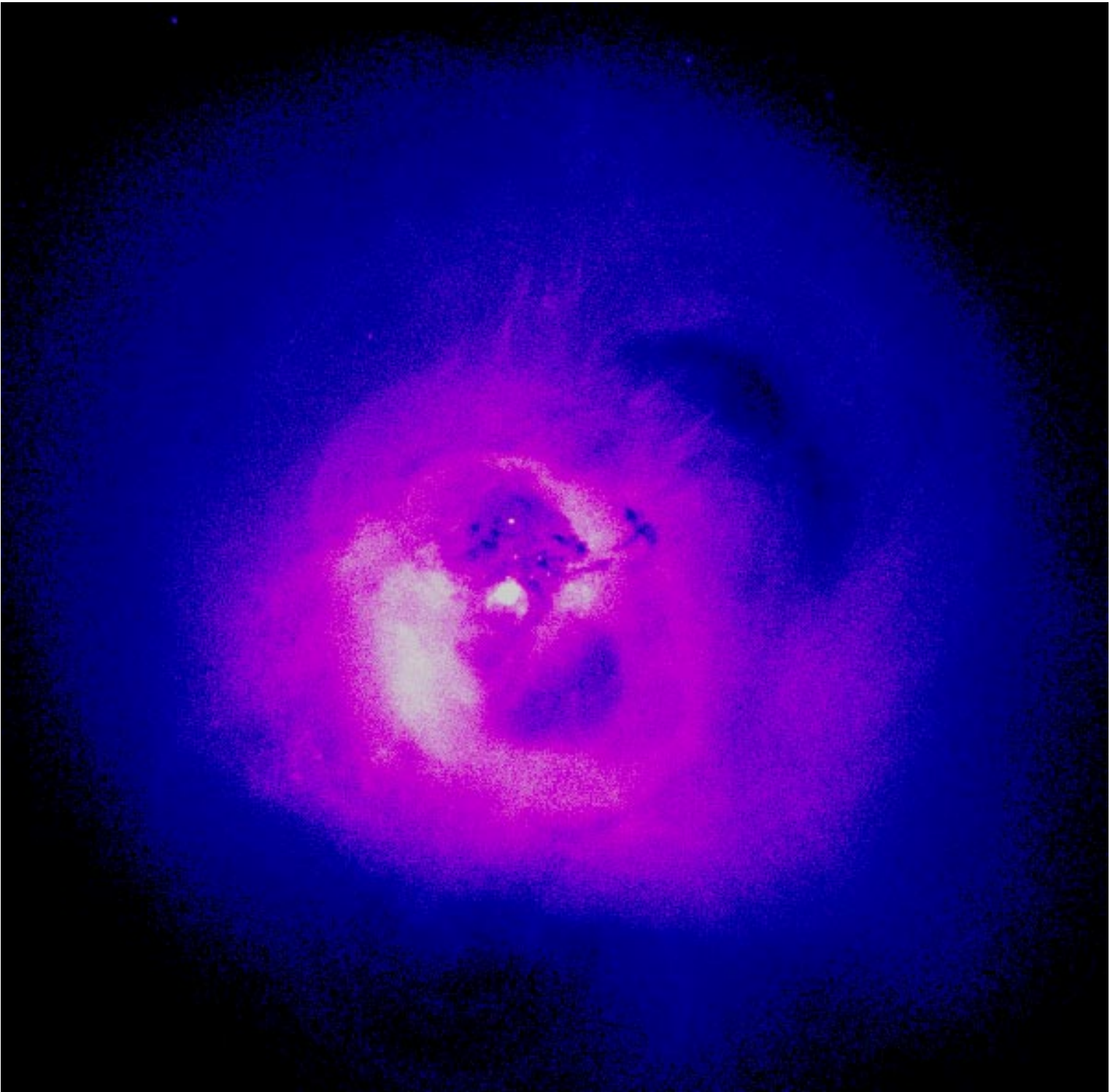}
\includegraphics[width=0.325\textwidth]%
{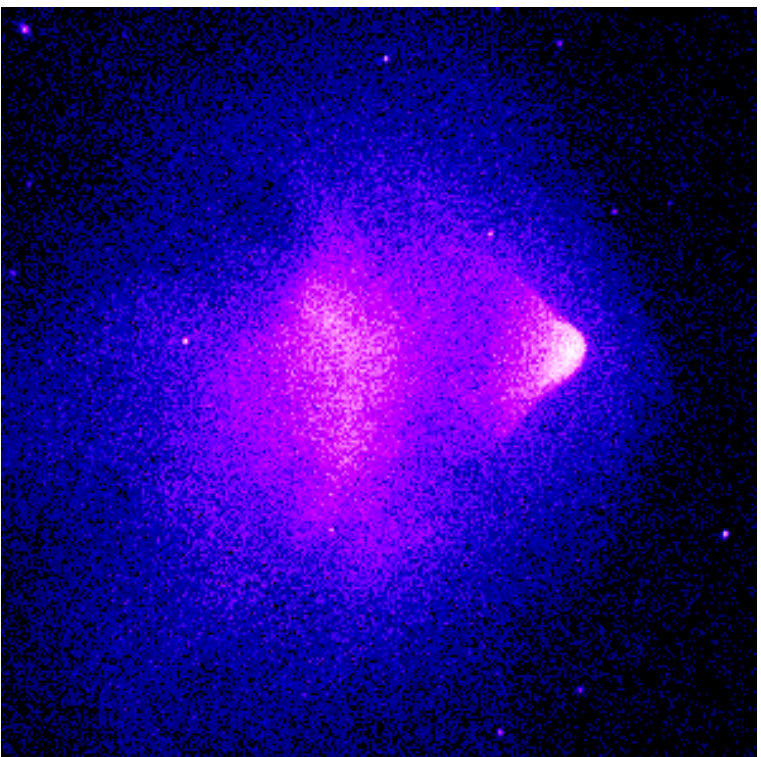}
\includegraphics[width=0.325\textwidth]%
{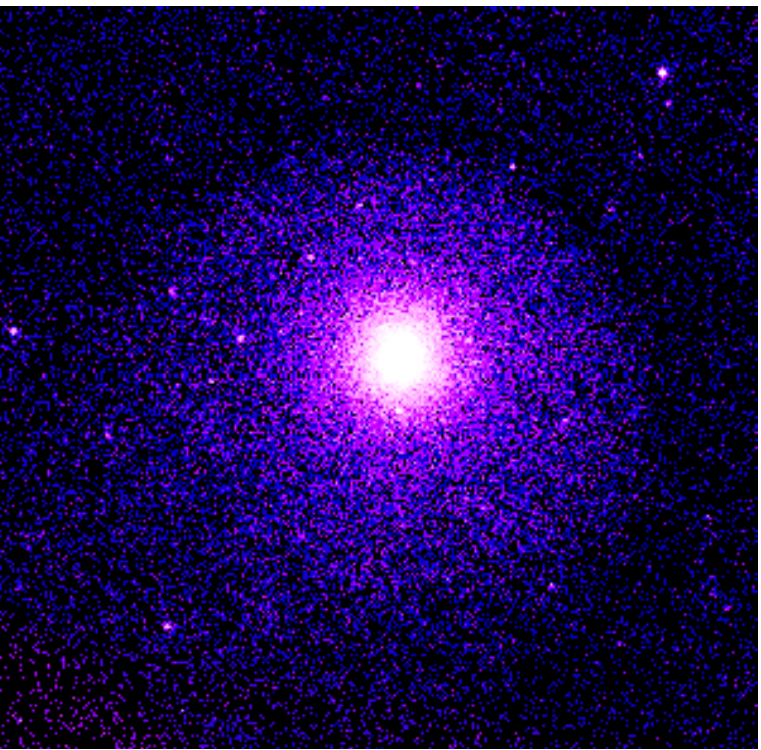}

\caption{\chandra\ X-ray images of three classes of high-velocity phenomena
  in intracluster plasma. Left: AGN bubbles and ripples in
  Perseus\cite{fabian03ripples}.  Middle: merger shock front and
  ram-pressure stripping in 1E0657-56\cite{m02textbook}. Right: a cold
  front apparently stabilized by magnetic draping in NGC\,1404 (a circular
  brightness edge spanning 180\deg\ from SW to NE)\cite{machacek05n1404}.
  Panels are $5'\times 5'$, approximately the FOV of the \genx\ 
  calorimeter.}

\label{fig:per}
\end{figure*}
%%%%%%%%%%%%%%%%%%%%%%%%%%%%%%%%%%%%%%%%%%%%%%

%%%%%%%%%%%%%%%%%%%%%%%%%%%%%%%%%%%%%%%%%%%%%%%%%%%%%%%%%%%%%%%%%%%%%%%%%%%
\section{CLUSTER PHYSICS}
\label{sec:phys}

Some of the questions posed in the cosmology section (\S\ref{sec:cosmol})
are interesting from the astrophysical perspective. Clusters, as well as hot
coronae of groups and massive elliptical galaxies, offer a unique laboratory
for plasma physics and hydrodynamics, with applications well beyond the
cluster field.

%%%%%%%%%%%%%%%%%%%%%%%%%%%%%%%%%%%%%%%%%%%%%%%%%%%%%%%%%%%%%%%%%%%%%%%%%%%
\subsection{Cluster tomography}
\label{sec:vels}

An imaging calorimeter is the next frontier of cluster astrophysics.
\astroh\ and \ixo\ will uncover large-scale gas flows and map the turbulence
in merging and relaxed clusters (WP by Arnaud et al). \ixo\ will determine
whether any turbulence is present in relaxed clusters, such as A1835 and
A2029, that are used for the $f_{\rm gas}$ cosmological test. It will also
directly test the hypothesis that turbulence is responsible for radio
halos\cite{brunetti08natur}. An imaging calorimeter such as \ixo\ can derive
the spectrum of turbulence and disentangle it from bulk flows by sampling
multiple, close lines of sight.

However, there are areas where the calorimetric spectral resolution needs to
be combined with arcsecond angular resolution.  These include sharp features
such as cluster shock fronts ({\em Is there turbulence behind the shock due
  to plasma instabilities? What is the post-shock ion temperature?}), cold
fronts and buoyant bubbles ({\em Why are they so stable?})  Resolving these
features will provide unique measurements of microphysical properties of the
intracluster plasma, such as viscosity, electron-ion equilibration and
ionization timescales, and the structure of the magnetic fields. These
properties govern mixing and transport processes and possibly heating of the
ICM by sound waves from the central black hole\cite{fabian03ripples}.  Fig.\ 
\ref{fig:per} shows examples of such objects --- the discovery potential of
calorimetric ``data cubes'' that map the ICM density, temperature and radial
velocity velocity with a super-\chandra\ angular resolution, is obvious.

%%%%%%%%%%%%%%%%%%%%%%%%%%%%%%%%%%%%%%%%%%%%%%%%%%%%%%%%%%%%%%%%%%%%%%%%%%%
\subsection{Resolving the Bondi radius}

Central AGN deposit enough energy into the dense, cool central gas in
clusters to prevent it from catastrophic
cooling\cite{mcnamaranulsen07cfrev}.  To accomplish this, but also avoid
blowing up the whole gas core, a feedback cycle is required to link cooling
of the hot gas to fueling of the AGN.  Understanding this process is of
fundamental importance.  A key part of a feedback cycle is the feeding of
the AGN.  The Bondi accretion model, which describes spherical, isentropic
accretion onto a black hole, can be radically modified in several ways.
Tiny angular momentum can reduce the accretion rate by orders of
magnitude.\cite{progabegelman03a,progabegelman03b} Dissipation of angular
momentum can affect the flow significantly\cite{blandfordbegelman99}. An
X-ray study of the gas flow in the region where it first comes under the
dominant influence of the black hole, i.e., near the Bondi radius, would
provide important clues to the factors that govern accretion of hot gas by
AGN.  However, for nearby supermassive black holes, including the important
case of M87, the Bondi radius is $\sim 1''$ (e.g., \citen{allen06bondi}),
requiring super-\chandra\ angular resolution.

%%%%%%%%%%%%%%%%%%%%%%%%%%%%%%%%%%%%%%%%%%%%%%%%%%%%%%%%%%%%%%%%%%%%%%%%%%%
\section{INTERGALACTIC MEDIUM}
\label{sec:whim}

Beyond the cluster infall shock lie giant filaments of the Cosmic Web,
filled with tenuous, warm-hot intergalactic medium (WHIM) that should
comprise more than half of the baryonic matter in the present-day Universe
\cite{miralda96whim,cenostriker99,cenostriker06}. The currently unseen
extended gas coronae of field galaxies (circumgalactic medium, CGM) may
contain an additional significant fraction. As a depository of all the gas
ever expelled from galaxies, this vast reservoir of baryons holds unique
information on the history of energy and metal production in the Universe.
By accurately accounting for these components, we can also directly
determine the present-day $\Omega_b$, to be contrasted with indirect
measures such as those from CMB.

%%%%%%%%%%%%%%%%%%%%%%%%%%%%%%%%%%%%%%%%%%%%%%%%%%%%%%%%%%%%%%%%%%%%%%%%%%%
\subsection{Absorption line studies}
\label{sec:whim-abs}

Unfortunately, this gas is very tenuous, which makes its X-ray and UV
emission extremely difficult, and for most of the WHIM, virtually impossible
to detect, which is why these ``missing baryons'' have never been seen.  At
temperatures $10^{5-6.5}$ K, they can be detected in absorption against
background beacons such as quasars and GRB X-ray afterglows.  However, the
absorbing columns are also very low.  Despite extensive observational
efforts, the vast majority of this dominant matter component remains
undetected.  FUV observations of O{\sc vi} and broad Ly$\alpha$ absorbers
\cite{danforthshull08,tripp08,richter06} probe only the lower-temperature
$10$\% of the missing baryons.  The rest should be searched for in the soft
X-ray band in absorption by highly ionized O, C, Ne, Fe.  Studies with
\chandra\ and \xmm\ came tantalizingly close to first
detections\cite{nicastro05whima,nicastro05whimn}, but much greater
sensitivity is required to discover the bulk of the missing baryons.

\ixo, as well as some proposed dedicated missions, will have the line
sensitivity factor%
\footnote{$S\equiv \sqrt{R A}$, where $R$ is spectral resolving
power and $A$\/ is the effective area.}%
10--20 times higher than \xmm\ and \chandra, and can be the discovery
facilities for WHIM studies. \ixo\ should detect at least 100 O{\sc vii}
absorbers over the sky (WP by Bregman et al).  Looking past \ixo, the
necessary next steps will be (a) to go much deeper in flux and sample the
Cosmic Web with thousands of sight lines, (b) to detect different absorption
lines of the same element in the same system and use line ratios to derive
the gas parameters, such as temperature and density, and (c) ultimately to
resolve those spectral lines, probing turbulence, disentangling it from
thermal broadening, and possibly determining ion temperatures. A large
number of detected absorbers will provide an independent dataset for
cosmological studies using the absorber cross-correlation analysis, yielding
constraints at redshifts and linear scales very different from those of the
CMB or Ly$\alpha$ forest.  Line positions and widths will allow us to study
the dynamics of WHIM. By separating different physical and kinematics phases
of the same WHIM system, one will be able to associate a given metal
component with HI systems seen in the UV (by future instruments), and thus
measure both the absolute and relative metallicity. This would give a direct
measure of the total baryon density $\Omega_b$.

The envisioned grating spectrometer on \genx\ will provide a further
increase in line sensitivity over \ixo\ by factor of 40, and a big increase
in spectral resolution, opening this nascent field of astronomy to truly
quantitative studies.

%%%%%%%%%%%%%%%%%%%%%%%%%%%%%%%%%%%%%%%%%%%%%%%%%%%%%%%%%%%%%%%%%%%%%%%%%%%
\subsection{X-raying the cluster outskirts}
\label{sec:clust-abs}

In addition to X-ray emission studies (\S\ref{sec:outskirts}), gas in the
cluster outskirts can be studied in absorption, just like the WHIM. The gas
between $r\sim 1-2\,r_{200}$ should have $T\sim 0.5-1$ keV and feature
prominent Fe and O absorption lines. It should also have higher column
densities than the more tenuous WHIM in the filaments. For \genx, several
hundred suitably bright background sources behind a $1-2\,r_{200}$ region of
the Coma cluster will be available for absorption line detection with
feasible exposures. Even a $z=0.2$ cluster such as A2163 should have $\sim
10$ suitable background quasars in the corresponding area.  This will open a
unique possibility to combine emission and absorption by the same medium and
directly observe, for example, the ionization nonequilibrium and stripping
of metal-rich gas from small halos in the cluster infall region.

%%%%%%%%%%%%%%%%%%%%%%%%%%%%%%%%%%%%%%%%%%%%%%%%%%%%%%%%%%%%%%%%%%%%%%%%%%%
\subsection{Using clusters as background sources}
\label{sec:clust-bg}

With a large effective area and an imaging calorimeter with a sufficient
FOV, it may be possible to use clusters as background candles for WHIM
absorption studies.  Clusters are among the brightest X-ray sources in the
sky and conveniently reside at the nodes of the Cosmic Web. Thus, cluster
sight lines are likely to pass through the densest regions of WHIM. An eV
spectral resolution is generally insufficient to detect low-column WHIM
systems (see WP by Bregman et al.), but column densities of some filaments
along the l.o.s.\ can be much higher\cite{m99sc}. They can produce
absorption lines blueshifted from the corresponding cluster emission lines.
Large angular extent of a cluster would result in volume-averaging over the
filament, providing highly complementary information to sparse, pencil-beam
sampling of the same filament by quasars.

%%%%%%%%%%%%%%%%%%%%%%%%%%%%%%%%%%%%%%%%%%%%%%%%%%%%%%%%%%%%%%%%%%%%%%%%%
\subsection{IGM in emission}
\label{sec:whim-emiss}

Recent numerical work suggests that metals may not be mixed well into the
bulk of the IGM, staying close to the field galaxies that produced them, in
the form of $100-200$ kpc plumes of the metal-rich galactic wind (e.g.,
\citen{bryan08cgm}). If so, these denser, more metal-rich clouds of
``circumgalactic medium'' (CGM) may just become detectable in emission.
Detecting and confirming their finite extent will provide critical input for
galaxy formation models and for the interpretation of the IGM absorption
line data.  CGM may also contain a significant fraction of the missing
baryons.

As outlined in \S\ref{sec:outskirts}, in addition to very large collecting
area, for such studies one will have to resolve the point-like and Galactic
diffuse components of the CXB using a calorimeter with a subarcsecond
angular resolution. In particular, CGM will emit a faint redshifted O{\sc
  vii} line that should be disentangled from a non-redshifted, but orders of
magnitude stronger, Galactic O{\sc vii} line.

Finally, a sensitive, low-background imaging calorimeter offers an exotic
possibility to observe WHIM in {\em reflection}, illuminated by bright
blazars whose beams are turned away from us, like a lighthouse in the fog
\cite{chur01blazarwhim}.

%%%%%%%%%%%%%%%%%%%%%%%%%%%%%%%%%%%%%%%%%%%%%%%%%%%%%%%%%%%%%%%%%%%%%%%%%
\section*{SUMMARY}

To summarize, we believe that the following fundamental questions in the
field of galaxy clusters and IGM will remain unanswered at the end of the
next decade:

$\bullet$ How much baryonic matter is hidden in the Cosmic Web, beyond the
confines of galaxies and clusters? What is the total amount and distribution
of heavy elements and thermal energy in this vast reservoir of baryons?

$\bullet$ What processes occur at the interface between the Cosmic Web
filaments and the galaxy clusters, and how do those processes affect cluster
physics and energetics?

$\bullet$ What hydrodynamic and plasma processes occur at shocks, cold
fronts and buoyant bubbles in clusters? How best to model them to use
clusters as a precision cosmology tool?

$\bullet$ How do supermassive black holes at the cluster centers feed on the
intracluster gas?

To answer these questions will require a futuristic X-ray facility that
combines a mirror with vast effective area and super-\chandra\ angular
resolution, a calorimeter array that matches the mirror resolution, and a
super high-dispersion grating.  Such a combination is envisioned as \generx.
For it to become a reality in our lifetime, key developments in the mirror
and detector technologies should occur in the coming decade (see \genx\ 
technical WP).  They build upon the \ixo\ technologies, but require a huge
leap forward from \ixo, thus depending critically on \ixo\ success.

% \section*{REFERENCES}
% 
% {\footnotesize http://hea-www.harvard.edu/$\sim$maxim/genx09/wprefs}

%%%%%%%%%%%%%%%%%%%%%%%%%%%%%%%%%%%%%%%%%%%%%%%%%%%%%%%%%%%%%%%%%%%%%%%%%%%
% \clearpage
% \bibliographystyle{prop}
% \bibliography{/home/maxim/TEX/PRO/GENX/genx_clusters_whitepaper09}

\end{document}